\documentclass[aps,prl,twocolumn,groupedaddress]{revtex4}
\usepackage{graphicx}
\usepackage{amsmath,amsfonts,amssymb}
\usepackage{color}
\usepackage{float}
\begin{document}
\title{Hydrogen-like Spectrum of Spontaneously
Created Brane Universes \\ with deSitter Ground State}

\author{Aharon Davidson}
\email{davidson@bgu.ac.il}
\affiliation{Physics Department, Ben-Gurion University
of the Negev, Beer-Sheva 84105, Israel}

\date{July 28, 2017}

\begin{abstract}
	Unification of Randall-Sundrum and Regge-Teitelboim brane
	cosmologies gives birth to a serendipitous Higgs-deSitter
	interplay.
	A localized Dvali-Gabadadze-Porrati scalar field, governed
	by a particular (analytically derived) double-well quartic
	potential, becomes a mandatory ingredient for supporting
	a deSitter brane universe.
	When upgraded to a general Higgs potential, the brane
	surface tension gets quantized, resembling a Hydrogen
	atom spectrum, with deSitter universe serving as the
	ground state.
	This reflects the local/global structure of the Euclidean
	manifold: From finite energy density no-boundary initial
	conditions, via a novel acceleration divide filter, to exact
	matching conditions at the exclusive nucleation point.
	Imaginary time periodicity comes as a bonus, with the
	associated Hawking temperature vanishing at the
	continuum limit.
	Upon spontaneous creation, while a finite number of
	levels describe universes dominated by a residual dark
	energy combined with damped matter oscillations, an
	infinite tower of excited levels undergo a Big Crunch.
\end{abstract}


\maketitle

\noindent\textbf{Introduction}

The no-boundary proposal \cite{NoBoundary} invokes
basic quantum mechanics to avoid the classically unavoidable
Big Bang singularity.
Creation in this language is a smooth Euclidean
to Lorentzian transition, with the emerging (finite scale
factor) universe resembling alpha decay.
The simplest model of this kind is constructed at the level
of the mini superspace, requires a positive cosmological
constant $\Lambda>0$,
and can only be implemented for a closed $k>0$ space.
A variant which introduces a supplementary
embryonic era can be realized, ad-hoc \cite{Vilenkin} by
including a radiation energy density term, field theoretically
by invoking the embedding approach \cite{Davidson},
or via the landscape of string theory \cite{Brustein}.
Brane extensions have also been discussed \cite{Hertog}.
The theoretical highlight of the no-boundary proposal is
the wave function of the universe, the solution of the 
Schrodinger Wheeler-deWitt (WdW) equation \cite{WdW}.

The two Randall-Sundrum (RS) models \cite{RS}, followed
by their Dvali-Gabadadze-Porrati (DGP) and Collins-Holdom
(CH) extensions \cite{DGP}
which supplement a 4-dim Einstein-Hilbert part to the
underlying 5-dim action, are presumably the prototype
brane models.
The first rights are reserved, however, to the Regge-Teitelboim
(RT) model \cite{RT} where the universe is treated as a 4-dim
extended test object floating geodesically \cite{GBG} in a 5-dim
non-dynamical background.
Moreover, the first field theoretically consistent brane variation,
albeit in a flat spacetime, was formulated by Dirac \cite{Dirac}.
Exporting the Dirac prescription to the gravitational regime
\cite{unified} allows us to treat the variety of models as special
limits of a single unified brane cosmology.
This Letter attempts to take the no-boundary proposal one
step further to expose the Hydrogen-like spectrum (with deSitter
as the ground state) of spontaneously created unified
brane universes.

\smallskip
\noindent\textbf{Unified brane cosmology in a nutshell}

Let the 4-dim FLRW cosmological
line element
\begin{equation}
	ds^{2}=-dt^{2}+a^{2}(t)
	\left( \textstyle{\frac{dr^2}{1-kr^2}}+r^2d\Omega^2 \right)
\end{equation}
be isometrically embedded within a $Z_2$-symmetric (L and
R branches, respectively) 5-dim AdS background characterized
by a negative cosmological constant $\Lambda_5<0$.
This can be done for any scale factor $a(t)$ and without
imposing any geometrical constraints.
The associated extrinsic curvatures are given explicitly by
\begin{equation}
	{\cal K}^{L,R}_{\mu\nu}=\left[
	\begin{array}{cc}
	\frac{1}{\xi}\left(\frac{\ddot{a}}{a}-
	\frac{1}{6}\Lambda_{5}\right)
	  & 0    \\
	0 & -\xi a^{2}\gamma_{ij}(r,\theta)
	\end{array}
	\right] ~.
\end{equation}
It is $\xi(a)$ which governs the cosmic evolution equation,
with the latter cast into the familiar FLRW format
\begin{equation}
	\frac{\dot{a}^2+k}{a^2}=\frac{\Lambda_5}{6}+\xi^2 (a) ~.
\end{equation}
Within the framework of unified brane cosmology \cite{unified},
in a nutshell, 
$\xi (a)$ is the root of the cubic equation
\begin{equation}
	\rho=\frac{3\xi^2}{8\pi G_4}
	+\frac{3\xi}{4\pi G_5}
	+\frac{\Lambda_5}{16\pi G_4}
	+\frac{\omega}{\sqrt{3}\xi a^4} ~.
	\label{cubic}
\end{equation}
$G_{4,5}$ denote the 4,5-dim gravitational constants,
respectively, and $\rho(a)$ stands for the localized DGP
energy density on the brane.
No specific equation of state $P=P(\rho)$ has been
assumed.
The $\omega$-term ($\omega$ is a conserved
charge),  resembles (but not to be confused with) the dark radiation
term which is known to accompany RS cosmology, is the
fingerprint of the underlying RT model.
It owes its existence to the built-in integrability of the
brane's geodesic equations of motion.
The special limits include:

\noindent$\bullet$ DGP limit ($\omega=0$):
The now quadratic eq.(\ref{cubic}) admits \cite{DGPcos}
two branches $\xi_{\pm}(a)$.

\noindent$\bullet$ RS limit ($\omega=0, G_4 \rightarrow \infty$):
$\xi_{+}(a)$ becomes proportional to $\rho (a)$, 
so that the FLRW
equation is unconventionally sourced \cite{RScos} by
$\rho_{total}=\frac{\Lambda_5}{2}+
\frac{1}{3}(4\pi G_5\rho)^2$.

\noindent$\bullet$ GR limit ($\omega=0, G_5 \rightarrow \infty$):
$\Lambda_5$ simply decouples.

\noindent$\bullet$ RT limit ($\omega \neq 0, G_5 \rightarrow \infty$):
The bulk is kept non-dynamical, $\Lambda_5 \neq 0$ is
optional.
Sticking to the original FLRW format, one formally
replaces $\rho$ by $\rho_{total}=\rho+\rho_d$,
compactly squeezing the entire deviation from
GR into an effective 'dark'  component
$\rho_d(\rho)$.
The latter must of course vanish for $\omega= 0$, obeying
\begin{equation}
	{\rho_d} ^2\left(8\pi G_4 (\rho+\rho_d)
	-\frac{\Lambda_5}{2}\right)
	=\frac{\omega^2}{a^8}~.
	\label{RTmaster}
\end{equation}

\noindent\medskip $\bullet$ In the general case
\cite{unified}, one may follow the formalism specified by eq.(\ref{RTmaster}),
only with modified $\{\rho^{\star},\rho^{\star}_d\}$
replacing $\{\rho,\rho_d\}$, where
$\rho^{\star}=\rho-3\xi/4\pi G_5$.

\medskip
\noindent\textbf{Higgs $\leftrightarrow$ deSitter interplay}
\smallskip

We start with a deceptively naive question:
\emph{What are the field theoretical ingredients necessary for
supporting a deSitter brane? }
It is well known that, within the framework of GR, introducing
a positive cosmological constant $\Lambda_4>0$ will do.
However, once a non-trivial $\rho_d(\rho)$ enters the
game, the answer is not straight forward any more.
Our goal is to end up with a constant $\xi(a)$.
Hence, the way to cancel out the $\omega$-term
in eq.(\ref{cubic}) is to arrange for a suitable energy density
\begin{equation}
	\rho(a)=\sigma+\frac{\omega}{a^4 \sqrt{\Lambda_4
	-\frac{1}{2}\Lambda_5}} ~.
\end{equation}
We are after a tenable field theoretical action capable of
(i) Sourcing the above radiation term,
(ii) Fixing the otherwise arbitrary $\omega$-charge, and
(iii) Bypassing fine tuning.
This can be achieved by introducing a DGP brane localized
real scalar field $\phi(x)$, subject to a particular uniquely
prescribed (analytically derived by means of reverse engineering
\cite{Higgs}) scalar potential $V(\phi)$.
 
The scalar potential and its derivative enter the game via the
intrinsic energy density
$V(\phi)=\rho-\frac{1}{2}\dot{\phi}^2$,
and via the Klein-Gordon (KG) equation
$V^{\prime}(\phi)=-\ddot{\phi}
-3\frac{\dot{a}}{a}\dot{\phi} $, respectively.
The constant value of
$\xi(a)^2=\frac{1}{3}(\Lambda_4-\frac{1}{2}\Lambda_5)$
then allows to express both of them as explicit functions
of $a$.
Self consistency of these two expressions is
achieved only provided a certain differential equation
is satisfied
\begin{equation}
	\frac{1}{4}{W^{\prime}(\phi)}^2=
	-k\sqrt{\frac{3}{\omega}\sqrt{\Lambda_4
	-\frac{1}{2}\Lambda_5}}
	W(\phi)^{\frac{3}{2}}
	+\frac{\Lambda_4}{3}W(\phi)~,
\end{equation}
where $W(\phi)=V(\phi)-\sigma$.
Counter intuitively, the exact analytic solution is surprisingly
familiar
\begin{equation}
	\boxed{V(\phi)=\sigma+
	\lambda^2 \left(\phi^2-v^2\right)^2}
	\label{V}
\end{equation}
A restricted Higgs potential has made its appearance
\begin{eqnarray}
	& \displaystyle{\lambda^2 =\frac{3k^2}{16\omega}
	 \sqrt{\Lambda_4-\frac{1}{2}\Lambda_5}~,~
	 \lambda^2v^2=\frac{\Lambda_4 }{12}}  ~, &
	 \label{lambdav} \\
	& \displaystyle{\sigma(\lambda, v)
	=\frac{\Lambda_4}{8\pi G_4}+
	\frac{\sqrt{3}}{4\pi G_5} 
	\sqrt{\Lambda_4-\frac{1}{2}\Lambda_5} } 
	\equiv \sigma_0~. & 
	\label{sigma0}
\end{eqnarray}
Note that the Higgs potential is a necessary but
not a sufficient ingredient for supporting a de-Sitter brane. 
Since $3 a^4\sqrt{\Lambda_4
-\frac{1}{2}\Lambda_5}\dot{\phi}^2=4\omega$,
the initial value $\dot{\phi}_c$, required by the
2nd order differential KG equation, gets fixed
by the initial scale factor value $a_c$.

The classical solution of the field equations is given by
\begin{equation}
	\textstyle{a(t)=\sqrt{\frac{3k}{\Lambda_4}}
	\cosh \sqrt{\frac{\Lambda_4}{3}}t ~, 
	\phi(t)=v
	\tanh \sqrt{\frac{\Lambda_4}{3}}t} ~.
\end{equation}
The symmetry role played by the so-called proper
scalar field $b(t)=a(t)\phi (t)/v$ is manifest via the
equilateral hyperbola $a(t)^2-b(t)^2=3k/\Lambda_4$.
Note that, in the spherically symmetric representation,
the deSitter spacetime is counter intuitively accompanied
by a \emph{non-singular} time dependent 'slinky'
($\phi\equiv 0$ on the horizon) scalar hair \cite{Higgs},
thereby avoiding the no-hair theorems of GR.

\medskip
\noindent\textbf{Variant Euclidization}
\smallskip

\begin{figure}[h]
	\center
	\includegraphics[scale=0.35]{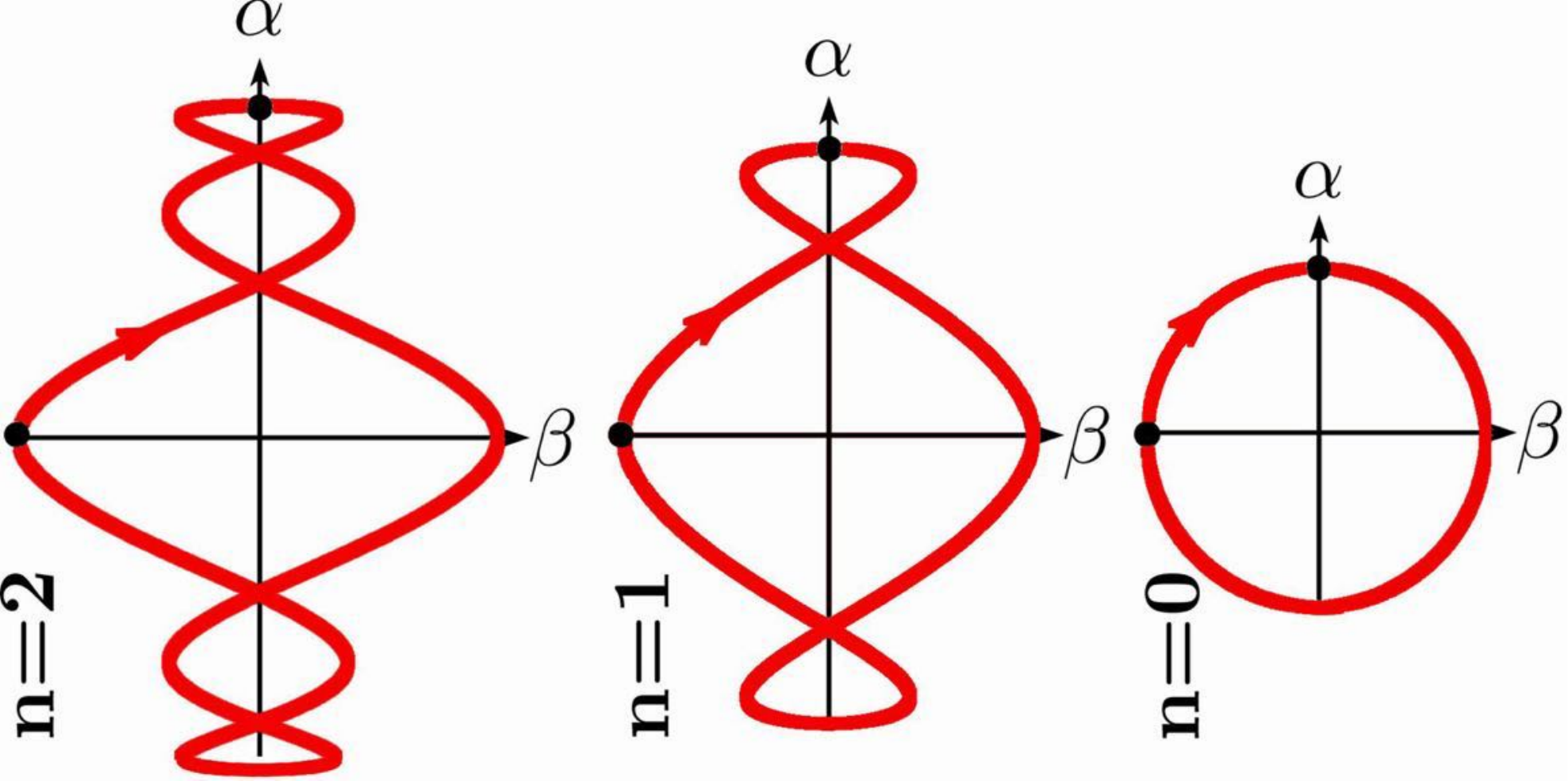}
	\caption{$\tau$-periodicity:
	Associated with the $n$-th level is a closed
	contour ($2n+1$ holes) in phase plane.
	The arrow leads from no-boundary (initial
	dot) to creation (final dot).}
	\label{Loops}
\end{figure}

Performing a Wick rotation $t\rightarrow i(\tau-\tau_c)$
implies
\begin{eqnarray}
	a(t) & \rightarrow & \alpha(\tau)
	=\textstyle{\sqrt{\frac{3k}{\Lambda_4}}
	\cos \sqrt{\frac{\Lambda_4}{3}}(\tau-\tau_c)}~, \\
	\phi(t) & \rightarrow & \textstyle{i\chi(\tau)
	=i v \tan \sqrt{\frac{\Lambda_4}{3}}(\tau-\tau_c)} ~,
\end{eqnarray}
normalized such that $\alpha(0)=0$, constituting
a circle in the Euclidean phase plane
$\alpha (\tau)^2+\beta (\tau)^2=3k/\Lambda_4$
(see Fig.\ref{Loops}), where $b(t)\rightarrow i\beta(\tau)$.
Globally, the de-Sitter imaginary time
periodicity $\Delta\tau=2\pi \sqrt{3/{\Lambda_4}}$
is now clearly manifest.
\begin{figure}[h]
	\center
	\includegraphics[scale=0.47]{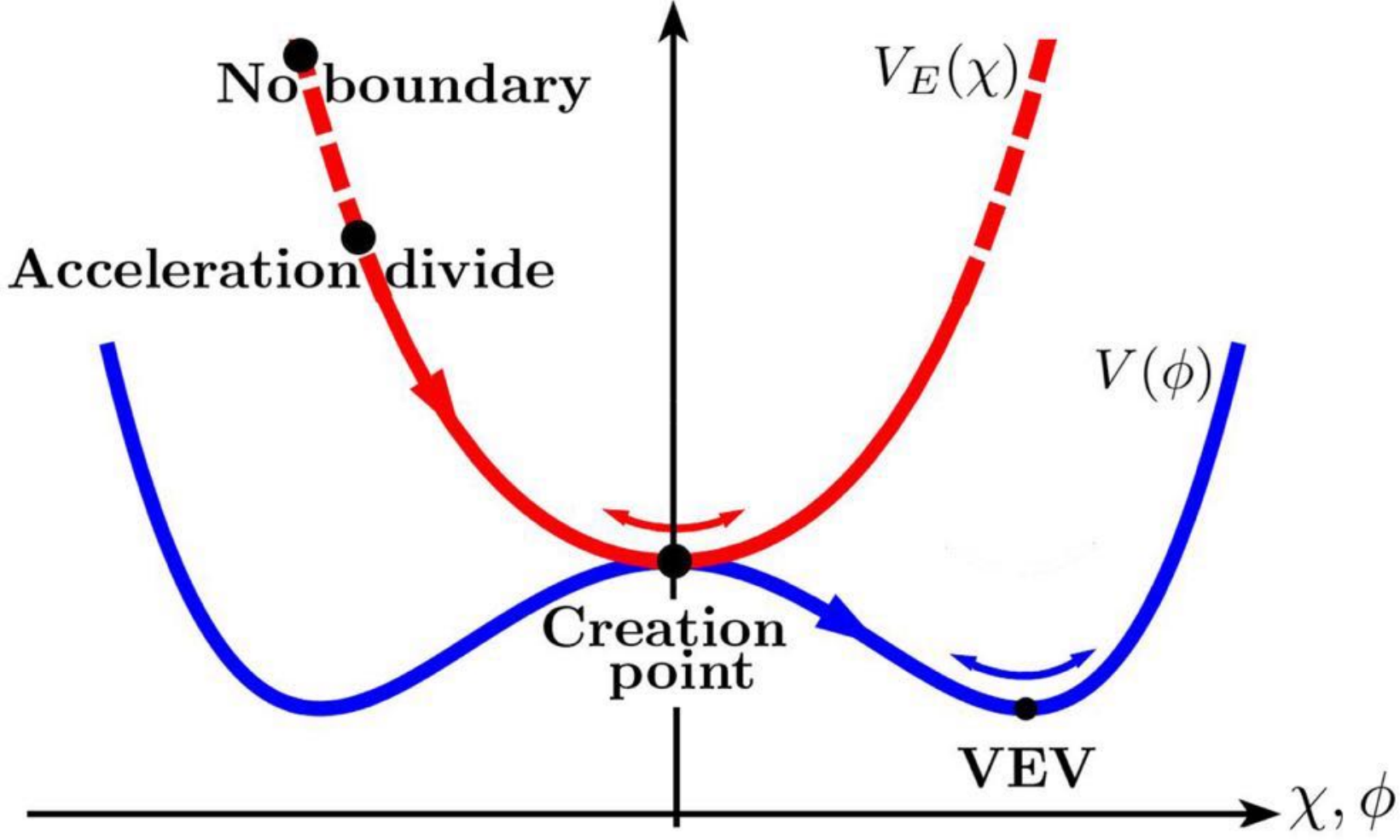}
	\caption{Euclide (red) $\leftrightarrow$ Lorentz
	(blue) transition takes place at the creation point.
	The number of $\chi(\tau)$ zeroes (passages
	through $V_E (\chi)$ minimum) per period
	is $2n$.
	Damped $\phi(t)$ oscillations around the VEV are
	interpreted as dark matter.}
	\label{Higgs}
\end{figure}

Had $a(t)\rightarrow \alpha (\tau)$ been conventionally
accompanied by $\phi(t)\rightarrow \chi (\tau)$, the
KG $\tau$-evolution in the Euclidean regime would
have been governed \cite{Colman} by $V_E (\chi)=-V(\chi)$.
This in turn would give rise to the well known
upside-down potential
$V_E (\chi)=-\sigma -\lambda^2(\chi^2-v^2)^2$.
However, in the present case
$a(t)\rightarrow \alpha(\tau)$ is unconventionally
accompanied by $\phi(t)\rightarrow i\chi (\tau)$
and hence $b(t)\rightarrow i\beta (\tau)$, so the
rules of the game are changed dramatically.
A closer inspection reveals that the KG equation,
while keeping its generic form, is actually being
governed by $V_E (\chi)=+V(i \chi)$, translated
in our case into
\begin{equation}
	\boxed{V_E (\chi)=\sigma +\lambda^2
	(\chi^2+v^2)^2}
	\label{VE}
\end{equation}
notably abandoning the upside down double well shape.

The scalar potential in the Lorentzian regime and
its companion in the Euclidean regime are given by
eqs.(\ref{V},\ref{VE}), respectively.
The two potentials touch each other in a single point.
The top of the $V(\phi)$ hill touches the
bottom of $V_E(\chi)$, see Fig.(\ref{Higgs}).
In turn, the quantum mechanical Euclidean to Lorentzian
transition, also known as creation, can exclusively take
place at
\begin{equation}
	\phi(0) = \chi_(\tau_c)=0 ~.
	\label{phimatch}
\end{equation}

\noindent\textbf{From no-boundary to spontaneous creation}
\smallskip

Allowing now for the most general (arbitrary
$\sigma,\lambda,v$) Higgs potential, the forthcoming
physics heavily depends on the local/global structure
of the Euclidean manifold.
We show that only a discrete class of smooth
Euclidean manifolds is in fact creative (= supports
Hubble-Hawking creation).
Our analysis stands on three legs:

\smallskip\textbf{(i) \textit{No-boundary:}}
The essence of the no-boundary proposal is keeping
the Euclidean origin perfectly smooth (singularity free).
The corresponding expansion reads
\begin{eqnarray}
	&& \alpha(\tau) \simeq \sqrt{k}\tau
	\left[1-\frac{\tau^2}{18}\left(
	\frac{256\lambda^4\omega^2}{9k^4}
	+\frac{\Lambda_5}{2}\right)\right],\\
	 && \beta(\tau) \simeq \frac{\sqrt{k}}{2\lambda v}
	\left[1-\frac{\tau^2}{18}\left(12\lambda^2v^2+
	\frac{512\lambda^4\omega^2}{9k^4}
	+\Lambda_5\right)\right].\hspace{25pt}
\end{eqnarray}
Consistently, as $\tau\rightarrow 0$, in spite of the
$\chi(\tau)\simeq 1/\lambda\tau$
behavior, the total energy density approaches
a finite value
\begin{equation}
	\rho_{total}\simeq\frac{\Lambda_5}{2}+
	\frac{256}{9k^4}\omega^2\lambda^4 ~.
\end{equation}
This regularity is rooted in the quartic term
of the Higgs potential.
The point to notice is that the entire $\{\sigma,\omega\}$
parameter space stays unconstrained at this stage.

\begin{figure}[h]
	\center
	\includegraphics[scale=0.45]{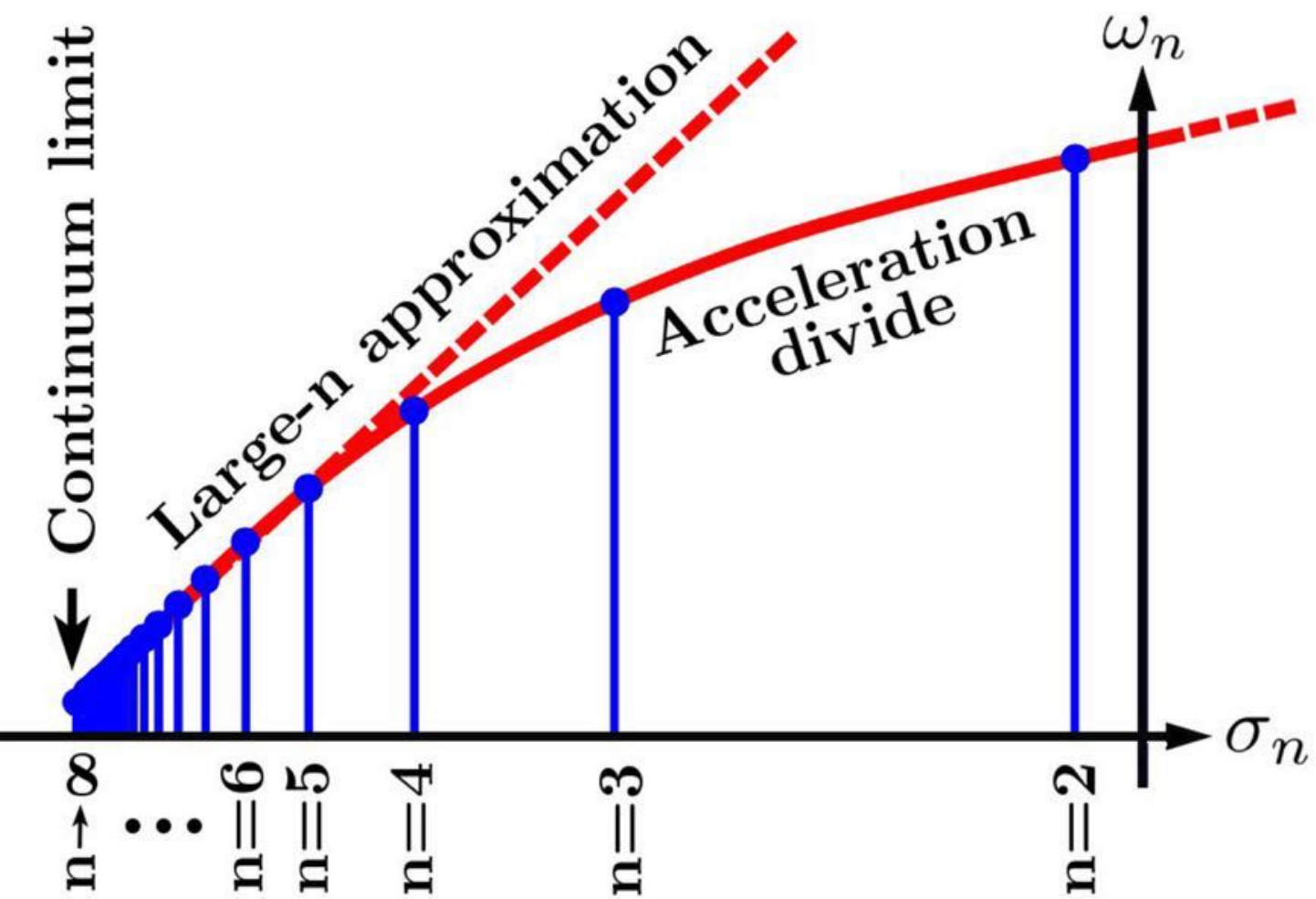}
	\caption{Hydrogen-like (large-$n$)
	surface tension spectrum (blue), demonstrated
	for the RT special case, is depicted along
	the acceleration divide curve $\omega(\sigma)$.
	deSitter ground state and the 1st excited
	state are located farther away on the rhs.}
	\label{Spectrum}
\end{figure}

\smallskip\textbf{(ii) \textit{Acceleration divide:}}
As the imaginary time advances, we arrive at a critical
crossroad where the cosmic acceleration, being of the
form $a^{\prime\prime}(\tau)=u(\tau)/d(\tau)$
gives rise to a novel $\omega$-selector which
resembles the Wien filter.
A zero of the $d(\tau)$ is potentially harmful, and
unless protected, is translated into an unacceptable
curvature singularity.
A zero of $u(\tau)$ humbly signals a velocity minimum.
Ironically, this is problematic from a different point of
view, as the creation point, characterized by
$a^{\prime}(\tau)=0$, does not really have a chance
to be reached (since once ${a^{\prime}(\tau)}$ starts
rising again, $d(\tau)$ turns positive definite).
The two zeroes must then cancel each other,
\emph{exactly as they do for the deSitter brane}.
This physical requirement can be algebraically formulated
by means two simple equations
\begin{equation}
	s(\tau_f)=1~,~~s^{\prime}(\tau_f)=0~,
\end{equation}
such that finite $a^{\prime\prime}(\tau_f)\neq 0$
(subscript f stands for 'filter').
The dimensionless function involved 
\begin{equation}
	s(\tau)=\frac{\frac{1}{3}\left[
	8\pi G_4 \rho-\frac{\Lambda_5}{2}
	+\frac{4G_4^2}{G_5^2}\right]^{\frac{1}{2}}}
	{\left[\frac{4\pi G_4 \omega}{3\sqrt{3} a^4}
	+\frac{G_4}{9G_5}\left(8\pi G_4 \rho-
	\frac{\Lambda_5}{2}   \right)
	+\frac{8G_4^3}{27 G_5^3}
	\right]^{\frac{1}{3}}} 
\end{equation}
constitutes the Vieta solution of the cubic eq.(\ref{cubic}).
In turn, for any value of the surface tension $\sigma$,
only a particular $\omega(\sigma)$, to be referred to
as the \emph{acceleration divide line}, remains physically
permissible.
The numerical solution is depicted in Fig.(\ref{Spectrum})
for the RT special case.

\smallskip
\textbf{(iii)\textit{Creation:}}
Equipped with the no-boundary initial conditions,
smoothly passing the acceleration divide filter with
a proper $\omega(\sigma)$ charge, the coupled
differential field equations are prematurely expected
to lead their solutions directly to the nucleation point.
This is, however, not necessarily the case as no
freedom is left to make every solution arrive at
\begin{equation}
	a^{\prime}(\tau_c)=0~,~~ \phi(\tau_c)=0~,
\end{equation}
at some finite $\tau_c$ (subscript c stands for 'creation').
In other words, not every Euclidean configuration
can undergo a quantum mechanical transition into a
Lorentzian universe.
Only a discrete spectrum of surface tensions
$\sigma_n  (n=0,1,2,..)$ will do.
Furthermore, associated with the surviving configurations
are already fixed $a(\tau_c)$ and $\phi^{\prime}(\tau_c)$
values.
No room for creation initial conditions.

\medskip
\noindent\textbf{Hydrogen-like spectrum} 
\smallskip

At this stage, we can offer exact analytic
$\{\sigma_n,\omega_n\}$ pair formulas for
the extreme levels $n=0$ and $n\rightarrow\infty$,
and semi-analytically extract (numerically verified) 
the leading $n$-dependence at the near continuum
(large $n$) approximation.
The $n=0$ ground state corresponds to the deSitter
brane.
Hence, eqs.(\ref{lambdav},\ref{sigma0}) apply.
It is by far the most isolated level.
For $n>0$, the scalar field $\chi(\tau)$ oscillates $n$
times around the bottom of the Euclidean potential
before finally reaching there the exact vanishing velocity
$\alpha^{\prime}(\tau_c)=0$ mandatory for creation.

The level density grows with n.
We refer to the highest $n\rightarrow\infty$ level
as the continuum limit.
The term 'continuum limit', coined to constitute an
analogy to quantum mechanics, is a bit misleading
though as it is only partially justified in the present
theory.
In fact, associated with this highest level is actually a phase
transition.
For $\sigma=\sigma_{\infty}+\epsilon^2$ (arbitrarily
tiny $\epsilon$), the velocity $\alpha(\tau)$ gradually
decreases, while the frustrated $\chi(\tau)$ spends
an infinite amount of imaginary time damped oscillating
around the bottom of the potential.
For $\sigma=\sigma_{\infty}-\epsilon^2$, on the other
hand, $\alpha(\tau)$ becomes a monotonically increasing
function of $\tau$, with $\chi(\tau)$ still damped oscillating,
and consequently, the solution totally loses its chance to
eventually hit the creation point.
At the continuum limit, the velocity gets stuck
$\alpha^{\prime}(\tau)\rightarrow \sqrt{k}$, and
$\chi(\tau)$ settles down at the minimum of the
potential, so that
$\rho(\tau)\rightarrow\sigma+\lambda^2 v^4$.
Altogether, the condition $\rho^{\star}=0$ (generalizing
$\rho=0$ in the presence of a finite $G_5$) is translated
into
\begin{equation}
	\sigma_{\infty}=
	\frac{3}{4\pi G_5}\sqrt{-\frac{\Lambda_5}{6}}
	-\lambda^2 v^4 ~.
\end{equation}

The full $\sigma$-spectrum, Hydrogen atom like for
large-$n$, numerically verified excellent
approximation, is given  by
\begin{equation}
	\sqrt{\frac{\sigma_0-\sigma_{\infty}}
	{\sigma_n-\sigma_{\infty}}}
	\simeq 2\left(n+1\right)-\frac{1}{n+1}~.
\end{equation}
This formula exhibits an empirical universal structure,
with the variety of parameters involved entering only
implicitly via $\sigma_{0}$ and $\sigma_{\infty}$.
The full $\{\sigma_n,\omega_n\}$ list, a collection
of points along the acceleration divide, is depicted in
Fig.\ref{Spectrum}
(for $\Lambda_5=0,G_5\rightarrow\infty$).
Note the linear large-$n$ approximation,
and notice the fact that $\omega_{\infty}\neq 0$.

\smallskip
\noindent\textbf{Imaginary time periodicity}
\smallskip

The three-leg foundation of the Euclidean manifold,
achieved by means $\sigma$-quantization, comes with
a bonus.
Namely, associated with every $n$-level there is a
loop in the $\{\alpha(\tau),\beta(\tau)\}$ phase
plane.
These loops, see Fig.(\ref{Loops}), form closed
Lissajous like trajectories, and differ from each other
by means of their total
number $2n+1$ of holes.
Most importantly, they dictate
$n$-dependent imaginary time periodicities
$\Delta_n=4\tau_c$.
And since $\Delta_n$ and
$\sqrt{\sigma_n-\sigma_{\infty}}$ have been
numerically observed to share exactly the same
functional behavior, the corresponding Hawking
temperatures are consequently given by
\begin{equation}
	\boxed{T_n =\frac{1}{\Delta_n}=
	\frac{\lambda v}{\pi}
	\sqrt{\frac{\sigma_n-\sigma_{\infty}}
	{\sigma_0-\sigma_{\infty}}}}
	\label{HawkingT}
\end{equation}
The temperature ranges from $T_0=\lambda v/\pi$
for the de-Sitter ground state down to
$T_{\infty}\rightarrow 0$ at the continuum limit.

\begin{figure}[h]
	\center
	\includegraphics[scale=0.47]{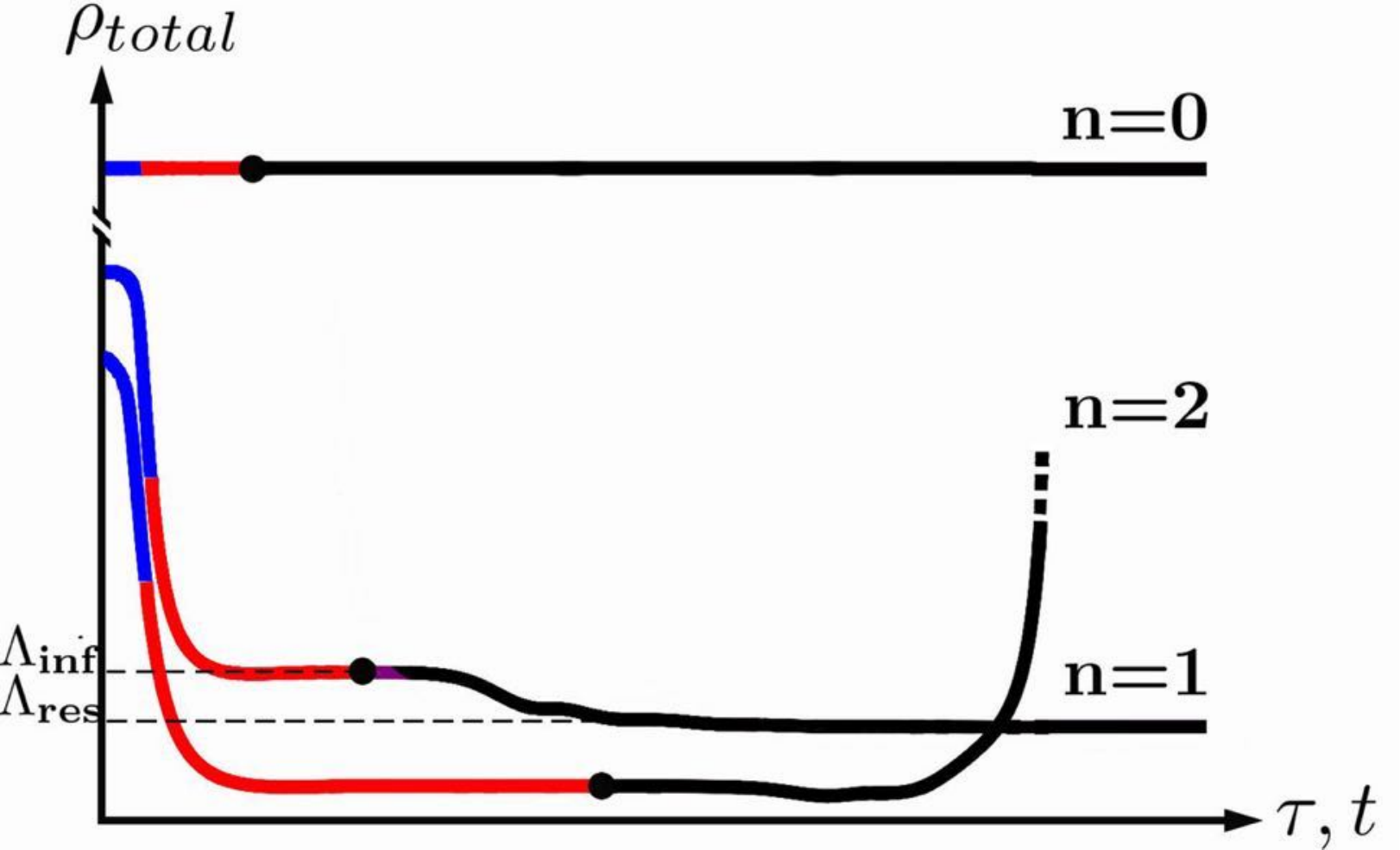}
	\caption{The total energy density evolution is
	depicted for the lowest $n$-states
	(here $\Lambda_5<0 $, $G_5$ finite).
	The black dots mark the Creation point.
	$n=0$ universe is deSitter.
	$n=1$ universe transforms from an inflationary
	$\Lambda_{inf}$ into a residual $\Lambda_{res}$CDM.
	$n\geq 2$ universes already suffer a Big Crunch.}
	\label{rhot}
\end{figure}

At the large-$n$ approximation, the Euclidean FLRW
$\tau$-evolution near the nucleation point is characterized
by an asymptotically constant $\rho_{total}(\tau)$, see
Fig.(\ref{rhot}).
On continuity grounds, the created Lorentzian universe is
then governed by the \emph{inflationary} cosmological
constant
\begin{equation}
	\Lambda_{inf}\simeq 12\lambda^2 v^2
	\frac{\sigma_n-\sigma_{\infty}}
	{\sigma_0-\sigma_{\infty}}>0~.
\end{equation}
As the time keeps ticking, the system is aiming toward
the local minimum of the Higgs potential associated
with the \emph{residual} DGP/RS/GR cosmological
constant
\begin{equation}
	\Lambda_{res}=\frac{\Lambda_5}{2}+
	3\left[\frac{G_4}{G_5}
	-\sqrt{\frac{G_4^2}{G_5^2}+
	 \frac{8\pi G_4}{3}\sigma_n-
	  \frac{\Lambda_5}{6}}
	 \right]^2~.
\end{equation}
Be aware, however, that in the present theory the
Higgs VEV is not always accessible.

\noindent The various options are demonstrated in Fig.\ref{rhot}:

\noindent$\bullet$ $n=0$ stands for the deSitter
universe.

\noindent$\bullet$ $n=1$ resembles $\Lambda_{res}$CDM,
and is somewhat closer to the universe that we observe.
The residual cosmological constant
$0<\Lambda_{res}\ll\Lambda_{inf}$ is accompanied
by dark matter (in average).
The damped oscillations of $\phi(t)$ near the VEV are
simply translated into ($p,q$ constants)
\begin{equation}
	\rho_{total}(t) \simeq \Lambda_{res}+
	\frac{p^2+q^2 \cos^2{\sqrt{8\lambda^2v^4
	-\frac{3}{4}\Lambda_{res}}t}}
	{\left(e^{\sqrt{\frac{\Lambda_{res}}{3}}t}\right)^3}~.
\end{equation}

\noindent$\bullet$ $n\geq 2$ eventually undergo
a Big Crunch.

\noindent In the most general case, a finite number
of states, characterized by
$4\pi G_5\sigma\geq \sqrt{-3\Lambda_5/2}$, is
associated with eternally expanding
$\Lambda_{res}>0$ universes.
This leaves an infinite tower of excited states
associated with eventually collapsing universes
(and hence $\Lambda_{res}=0$, no fine tuning).

\medskip
\noindent\textbf{Epilogue}
\smallskip

We have shown that associated with spontaneous
creation of a unified brane universe is a Hydrogen atom
like spectrum, with deSitter universe playing the role of
the ground state.
Our discussion has been carried out at the semi classical
level, with the focus on the local/global structure of the
4-dim Euclidean manifold.
Smoothness and regularity of the no-boundary Euclidean
manifold combined with the exclusiveness of the spontaneous
creation mechanism is bonused by imaginary time
periodicity.
In the near future we would like to
(i) Solve the WdW equation for the Hartle-Hawking
wave eigenfunctions,
(ii) Replace the present Bohr-like approach by a path
integral calculation,
(iii) Calculate the finite amplitude for creating a
universe similar to ours from nothing, and
(iv) Check applicability to the multiverse paradigm.
We would also like to mathematically reveal the
underlying black hole connection.
The generic imaginary time periodicity may hint
towards an infinite tower (so far, only the $n$=0 case has
been revealed \cite{Higgs}) of asymptotically cosmological
black hole configurations accompanied by slinky time-dependent
scalar hair.
And last but not least, elevating the Higgs-deSitter
interplay to a fundamental level, we dare to conjecture
potential relevance to GUT/Planck physics.

\acknowledgments
{Many thanks to Judy Kupferman for the critical reading and
helpful advice in organizing the paper.}

\end{document}